# Fluid Mechanics Explains Cosmology, Dark Matter, Dark Energy, and Life⋆

**Carl H. Gibson**

University of California at San Diego, Departments of Mechanical and Aerospace Engineering and Scripps Institution of Oceanography, Center for Astrophysics and Space Sciences, 9500 Gilman Drive, La Jolla CA 92093-0411, cgibson@ucsd.edu

**Abstract**

Observations of the interstellar medium by the Herschel, Planck etc. infrared satellites throw doubt on standard ΛCDMHC cosmological processes to form gravitational structures. According to the Hydro-Gravitational-Dynamics (HGD) cosmology of Gibson (1996), and the quasar microlensing observations of Schild (1996), the dark matter of galaxies consists of Proto-Globular-star-Cluster (PGC) clumps of Earth-mass primordial gas planets in metastable equilibrium since PGCs began star production at 0.3 Myr by planet mergers. Dark energy and the accelerating expansion of the universe inferred from SuperNovae Ia are systematic dimming errors produced as frozen gas dark matter planets evaporate to form stars. Collisionless cold dark matter that clumps and hierarchically clusters does not exist. Clumps of PGCs began diffusion from the Milky Way Proto-Galaxy upon freezing at 14 Myr to give the Magellanic Clouds and the faint dwarf galaxies of the $10^{22}$ m diameter baryonic dark matter Galaxy halo. The first stars persist as old globular star clusters (OGCs). Water oceans and the biological big bang occurred at 2-8 Myr. Life inevitably formed and evolved in the cosmological primordial organic soup provided by $10^{80}$ big bang planets and their hot oceans as they gently merged to form larger binary planets and small binary stars.

## 1. Introduction

Several simultaneous scientific revolutions are underway that affect cosmology. Turbulence has been narrowly defined based on the inertial-vortex force. This reverses the standard large to small Taylor-Reynolds-Lumley turbulence cascade direction and explains heat-mass-momentum-information transport of natural fluids in terms of fossil turbulence and fossil turbulence waves. From the new turbulence, the big bang is simply a Planck-Kerr instability similar to supernova pair production, but with sufficient temperature to extract mass-energy from the vacuum and produce the universe in an inflationary event. Anti-gravitational forces are supplied by vortex line stretching so that a permanent dark energy Λ is unnecessary. Big bang events are fossilized as turbulent length scales exceed horizon scales $L_H = ct$, where c is the speed of light and t is time. Gluon-viscous stresses power inflation by turbulent combustion, limited locally by vortex dynamics and terminated by thermodynamics. Gravitational structures begin during the plasma epoch at large photon-viscous scales with weak turbulence, not at small Cold Dark Matter CDM scales, as shown in Figure 1. Collisional fluid mechanics of HGD requires life to begin early and be spread rapidly and widely on cosmic scales. Intelligent extraterrestrial life is inevitable and everywhere according to HGD cosmology, but is virtually impossible from ΛCDMHC cosmology.

---

⋆ Presented 10 Aug. 2012 at NASA-AMES Symposium: A Universe of Scales: from Nanotechnology to Cosmology, to honor the memory of Minoru Freund.

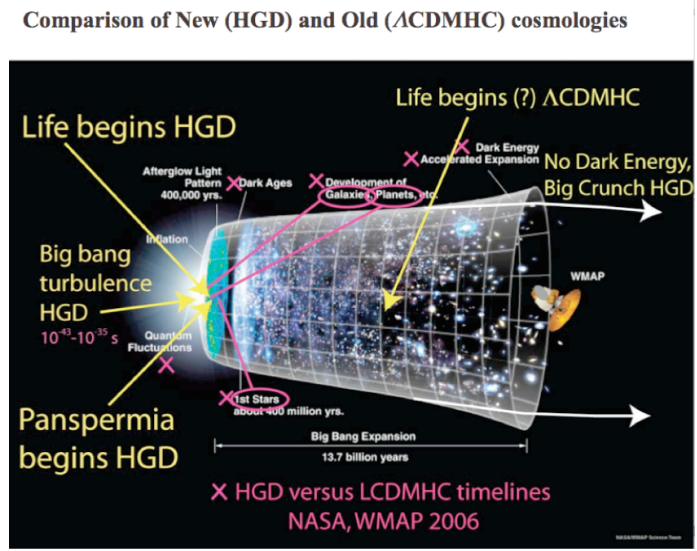

Figure 1. Time-line of New and Old cosmologies. Both require a cosmological big bang event at time zero. New cosmology explains the big bang as a turbulent combustion instability at $10^{32}$ K Planck temperature, where a Planck particle pairs with a Planck anti-particle and symmetry is broken by a prograde capture of a Planck anti-particle, Gibson (2004, 2005). Fossils of big bang turbulence trigger HGD proto-super-galaxy-cluster-void fragmentation at 30,000 years, not 14 Gyr according to Old cosmology. The viscous fragmentation scale at plasma to gas transition changes from galaxy to planetary mass. The planets are in Jeans mass clumps with the mass of a globular star cluster. These $10^{80}$ planets become the dark matter of all galaxies and the source of all stars, Gibson (1996), Schild (1996), and host the formation of water oceans and life at 2 million years (the biological big bang), not 400 million years at the end of the mythical dark ages. Failures of Old cosmology are shown by X symbols. ΛCDMHC cosmology predicts the big bang and inflation events, but fails to describe HGD gravitational structure and life formations.

Hierarchical HC clustering of CDM plasma seeds is fluid mechanically untenable and contrary to observations. Forget ΛCDMHC cosmology. It doesn't work. Instead, fossil remnants of big bang turbulence trigger fragmentation of the ex-panding plasma when decreasing photon-viscous force scales $L_{SV}$ are matched by increasing scales of causal connection $L_H$. The resulting mass scale is that of a super-cluster of galaxies, or $10^{46}$ kg, and the time is $10^{12}$ seconds, or 30,000 years, after the cosmological big bang. The mass scale of fragmentation decreases to that of proto-galaxies ($10^{43}$ kg) during the plasma epoch until $10^{13}$ seconds (300,000 years) when the hydrogen-helium plasma becomes gas. The kinematic viscosity decreases by $10^{13}$, reducing the fragmentation mass to $10^{25}$ kg, the mass of the Earth. A second fragmentation scale is that of Jeans 1902, $L_J = V_S \tau_G$, where $V_S$ is the speed of sound, $\tau_G$ is = $(\rho G)^{-1/2}$ the gravitational free-fall time, and $\rho$ is the density. The density of globular star clusters in the Milky Way and other galaxies is observed to match $\rho_0$ of the plasma at the time of first fragmentation ($4 \times 10^{-17}$ kg m$^{-3}$), which is $4 \times 10^4$ times larger than the density of galaxies $\rho_{MilkyWay}$.

As shown in Fig. 1, numerous cornerstones of Old cosmology vanish when modern fluid mechanics is applied. Stars form from a binary cascade of merging planets. Stars are therefore generally observed as binary star systems. The constant recycling by reducing hydrogen-atmosphere planet-mergers seeded by chemicals produced and spread as oxides by dying stars is reflected in the ample abundance of water, exo-planets, and organic waste-products of extraterrestrial life, Wickramasinghe et al. (1997), Pflug and Heinz (1997), Hoover (2011). Interstellar dust is mostly polycyclic aromatic hydrocarbons (oils), not ceramics, produced by the barbecue smoke of life-infested-planet mergers. Millions of planets exist per star in all galaxies, not the handful expected by the standard cosmological model where stars make all the planets and their metal cores and deep water oceans are mysteries.

The first stars were small and long-lived, as found in old globular clusters, not short-lived massive superstars that re-ionized the universe. Re-ionization never happened. Temperatures at the time of first stars for Old cosmology were so low that life formation would have been rare, local, and virtually impossible. Rather than accelerating expansion of a dark energy dominated universe driven by antigravity, the HGD universe revealed by fluid mechanics must be more dense than the critical density of a flat universe ($10^{-26}$ kg m$^{-3}$), and will end in a big crunch. The 2011 Nobel Prize in Physics is falsified by the existence of life, Gibon (2011).

Numerous space telescopes, especially in the infrared frequency bands pioneered by Minoru Freund and his Japanese colleagues in their 1995 measurements from the Infrared Telescope in Space IRTS, provide a flood of detailed information about the PGC clumps of frozen dark matter planets. Figure 2 (Gibson 2012, Fig. 3) shows an example of such information from the Herschel space observatory and Planck infrared satellites, both launched simultaneously on May 14, 2009, to orbits near the Earth-Sun second Lagrange point. Each of the red dots in the upper left image is interpreted here as a PGC proto-globular-star-cluster. The red dots of Fig. 2 (upper left) are presented (ESA/NASA/JPL-Caltech/STScI) as ceramic dust clouds. The high resolution image (lower right) of a closer PGC confirms they are Jeans mass clumps of frozen primordial gas planets, each with enough material to form a globular star cluster of a million solar mass stars.

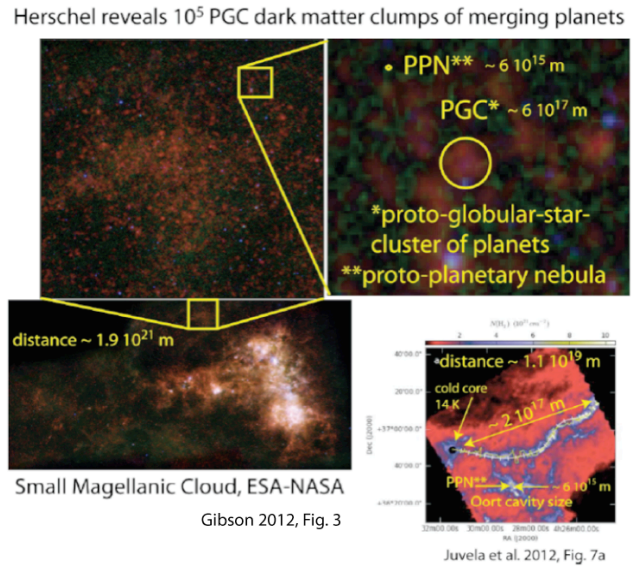

Figure 2. High resolution Herschel space observatory images reveal the dark matter PGC clumps of mostly frozen earth-mass planets that make the stars of the Small Magellanic Cloud, top and left. Each of the red dots shown at top left can be identified as a proto-globular-star-cluster PGC containing a trillion planets, but are undetectable at visible frequencies. The interior of a much closer PGC is shown at lower right, with a Cold Core (14 K) object and its filament from Planck. From its size ($6 \times 10^{15}$ m) and temperature (14 K) the Cold Cores and filament widths are interpreted as Oort Cavity sizes, where an Oort Cavity is the size $L_{OC} = (2M_{SUN}/\rho_0)^{1/3}$ left by a pair of stars when sufficient planets of a PGC assemble by mergers. The mergers are triggered by the center of gravity of another PGC drifting through the nearby PGC examined, at a distance of $1.1 \times 10^{19}$ m, leaving an $L_{OC}$ wide filament of length $2 \times 10^{17}$ m as it crosses the PGC.

The 14 K temperature of cold cores catalogued by the Planck observations reflects the 13.8 K triple point of hydrogen as planets evaporate and merge to form stars. Planet clouds can persist for billions of years in metastable equilibrium, but ceramic dust clouds are gravitationally unstable and would promptly collapse.

## 2. Theory

Turbulence is defined as an eddy-like state of fluid motion, where the inertial-vortex forces of the eddies are larger than any other forces that tend to damp the eddies out, Gibson (1980, 1996). The conventional definition of turbulence is less specific. Turbulence syndrome lists (Tennekes and Lumley, Pope) presented like the symptoms of a disease (e.g.: diffusive, chaotic, non-laminar, etc.) fail to distinguish active from fossil turbulence. New turbulence provides hydrodynamic phase diagrams, Gibson (1980). Old turbulence assumes turbulent kinetic energy cascades from large scales to small, Taylor (1938). The opposite is true.

Fossil turbulence is defined as a perturbation in any hydrophysical field produced by turbulence that persists after the fluid is no longer turbulent at the scale of the perturbation. Inertial vortex forces per unit mass **v**×**ω** arise in the Navier Stokes equations that express the conservation of momentum for a fluid, where **v** is the velocity and **ω** is the vorticity of the fluid. Turbulence always forms at the Kolmogorov scale $L_K = (\nu^3/\varepsilon)^{1/4}$, where $\nu$ is the kinematic viscosity and $\varepsilon$ is the viscous dissipation rate of kinetic energy per unit mass.

The mechanism of the turbulence cascade is **v**×**ω** driven merging of adjacent eddies with the same spin direction to form larger eddies. The turbulence cascade proceeds from small scales to large by such eddy mergers. To understand turbulence, the Navier-Stokes equations should be written

$$\partial \vec{v}/\partial t = -\nabla B + \vec{v} \times \vec{\omega} + \vec{F}_{viscous} + \vec{F}_{buoyancy} + \vec{F}_{Coriolis} + \vec{F}_{other} \quad (1)$$

so that the non-linear inertial vortex force term is isolated from the Bernoulli group of energy terms. In equation (1), the Bernoulli group is

$$B = v^2/2 + p/\rho + lw$$

where the combination B is often constant in turbulent flows so its gradient $-\nabla B$ disappears in the momentum equation (1). The first B term $v^2/2$ is the kinetic energy per unit mass, the second term $p/\rho$ is the specific stagnation enthalpy, and the last term $lw$ is the specific lost work due to viscous dissipation. Dimensionless groups are used to determine whether turbulence is possible. The Reynolds number $Re = \vec{v} \times \vec{\omega} / \vec{F}_{viscous}$ is the ratio of the inertial-vortex force to the viscous force. The Froude number is the ratio of the inertial-vortex force to the buoyancy force, etc. By this definition, flows are not turbulent if their Reynolds numbers, Froude numbers, Rossby numbers, etc. are not larger than universal critical values.

Only when **v**×**ω** dominates all other terms in the momentum equation should the flow be considered turbulent. Kolmogorov-Obukhov universal similarity hypotheses for turbulence can be understood as a cascade from small scales to large driven by **v**×**ω** vortex dynamics.

Figure 3 shows the HGD cosmology theory using our new turbulence definition and cascade direction to describe the big bang cosmological event and viscous inflation, from Gibson, Schild and Wickramasinghe (2011), Fig. 8. Ideal flow is assumed in conventional applications of the general relativity theory, but this is inappropriate under Planck conditions where Planck temperatures of $10^{32}$ K only permit Planck mass particles and anti-particles to exist, with mean free paths for collision equal to the Planck length scale $10^{-35}$ m, Gibson (2004, 2005). Dimensional analysis shows the Kolmogorov length scale matches the Planck scale. Planck-Kerr instability develops where a Planck particle and a Planck antiparticle pair in a stable mode similar to positronium formation from an electron and a posi-tron at super-nova temperatures of $10^{10}$ K. Prograde accretion of a Planck antiparticle releases 42% of the rest mass energy, making the big bang the first turbulent combustion, since all this energy must go to the production of more Planck particle pairs. Very little entropy is produced at such high temperatures. However, open and flat universes are ruled out by big bang entropy from turbulence dissipation.

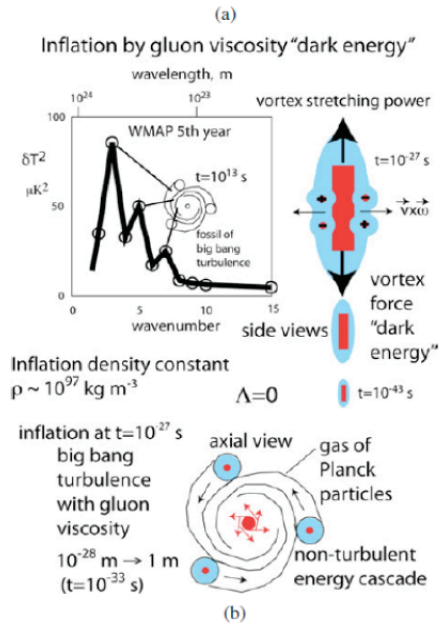

Figure 3. Cascade of big-bang turbulence from Planck scales to strong force freeze-out scales where quarks form and negative gluon viscous stresses drive inflation to meter scales at Planck densities. Inertial-vortex forces exceed $10^{113}$ Pa, extracting mass-energy from the vacuum by stretching vortices generated at Planck scales in a non-turbulent energy cascade from large scales to small. Anti-gravitation vortex forces render a permanent dark energy $\Lambda$ unnecessary. The largest scale WMAP signals between $10^{24}$ and $10^{23}$ m reflect the primary and secondary vortices expected for the turbulence event.

Temperature and vorticity fluctuations are fossilized by the inflation event which produces velocities much larger than the speed of light and are stretched to much larger scales as the universe and space expands. Structure formation begins again when the scale of causal connection ct increases with time t until it matches the viscous Schwarz scale $L_{SV} = (\gamma\nu/\rho G)^{1/2}$, where $\gamma$ is the rate-of-strain, $\nu$ is the kinematic viscosity, G is Newton's gravitational constant, and viscous and self-gravitational forces are equal. This occurs during the plasma epoch at time $10^{12}$ seconds (30,000 years).

Fossils of the big bang turbulence event include vortex lines; that is, fossil vorticity turbulence. Our local spin vector is reflected in a preferred spin direction, now expanded to $10^{25}$ m scales by the expansion of the universe, Schild and Gibson (2008). Polarization effects of the spin may soon be detected in cosmic microwave background anisotropy signals, Keating et al. (2011).

### 3. Observations

As shown in Fig. 2, the Oort Cloud length scale $L_{OC}$ of planetary condensation to form stars is $6 \times 10^{15}$ meters, computed from the density $\rho_0$ of the plasma at the time of first fragmentation and the mass of a pair of stars. When such stars appear they can be expected to form a halo of partially evaporated frozen gas planets termed a proto-planetary-nebula with the same intrinsic length scale, labeled PPN in Fig. 2. Figure 4 shows a Spitzer infrared image of the Helix Planetary Nebula, perhaps the closest one to the Earth at a distance of $6.7 \times 10^{18}$ meters, about half the distance to the PGC shown in the lower right of Fig. 2.

Considering the large number of planets per star in a galaxy required for these to constitute the dark matter ($3 \times 10^7$), why have they not been observed? Pfenniger and Combes (1994) suggest the dark matter of galaxies should be cold gas cloudlets, but could not make a convincing argument for star formation using the Jeans 1902 instability. Walker, M. and M. Wardle (1998) detect extreme scattering events from Jupiter mass gas clouds with sufficient probability to permit the claim that they might contribute a significant fraction of the dark matter of the Milky Way Galaxy. Ceccarelli and Dominik (2006) predict the ortho-$H_2D^+$ 372 GHz line would permit detection of such cloudlets. Heithausen (2004) detected the cloudlets expected using carbon monoxide CO at a distance of $3 \times 10^{18}$ m with diameters a few times the size of the evaporated planets $10^{13}$ m shown in Fig. 4.

Microlensing studies EROS (Expérience de Recherche d'Objets Sombres) and MACHO (Massive-Compact-Halo-Objects) have examined light curves of millions of Magellanic Cloud stars nightly for years, and reject the possibility that >20% of the dark matter of the Galaxy might be in the Mars-Jupiter planetary-mass range, Alcock et al. (2001), contrary to Gibson (1996) and Schild (1996) who claim ~ 99%. EROS and MACHO collaborations assume wrongly that the planets are uniformly distributed and not clumped as PGCs of a trillion earth-mass frozen hydrogen planets, as we see in Fig. 2. Furthermore, the density of the dark matter planets within star forming clumps is highly non-uniform and the clumps are clumped in the Galaxy halo. All of these clumping factors cause mass function underestimates in such dark matter studies.

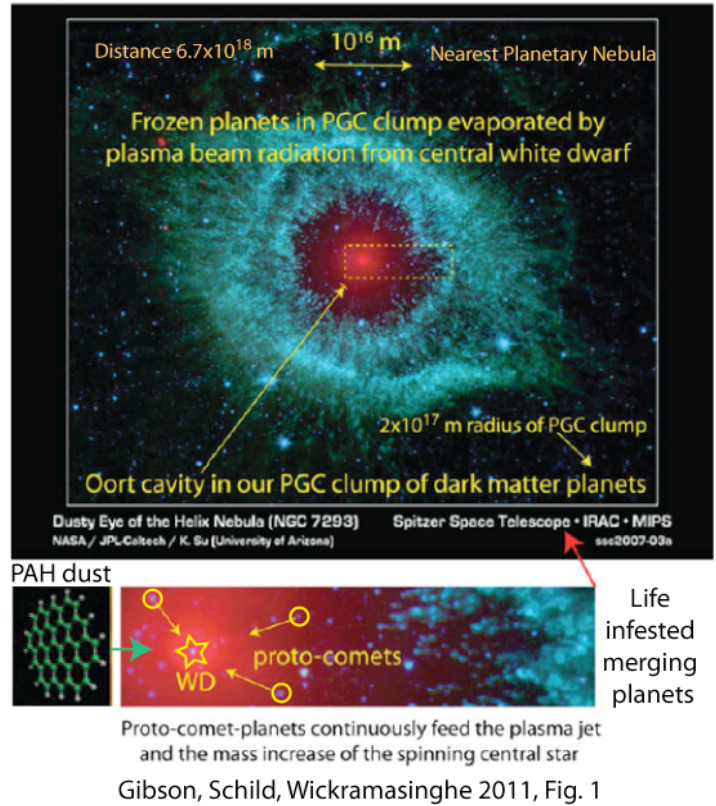

Figure 4. Helix planetary nebula shown by the infrared telescope Spitzer, from Gibson, Schild and Wickramasinghe 2011, Fig. 1. Frozen dark matter planets are partially evaporated by jets from the central white dwarf star WD, producing the Oort Cavity. At the bottom of Fig. 4 we see proto-comets moving toward the central star leaving long hydrogen rich wakes, feeding the WD mass toward the theoretical limit of 1.44 solar where a super-nova of the carbon star will occur. Planetary atmospheres at the Oort cavity boundary are often $10^{13}$ m in diameter, and could easily dim the light of a supernova by turbulent refractive index scattering if it were on the line of sight, producing a systematic dimming error and a false indication of accelerating expansion of the universe driven by anti-gravitational forces of dark energy $\Lambda$. As shown in Fig. 3, a permanent dark energy $\Lambda$ is unlikely and unnecessary. The red dust in the center of the image is composed of polycyclic-aromatic-hydrocarbons PAH molecules, strongly indicating the merged planets are life infested.

Further evidence of dark matter planets making stars in a PGC is shown in Figure 5. Because Herschel and Planck detect so many infrared frequency bands they can measure the temperature precisely to examine and catalog Planck cold cores within Planck cold clouds (Juvela et al. 2012 terminology). The pattern of radia-tion and temperature leaves little doubt that the objects studied are not dust clouds but PGC planet clouds with PPN proto-planetary-nebula triggered at Oort cavity scales on trajectories reflecting other PGC centers of gravity, as shown by dashed lines in Fig. 5 and the filament in Fig. 2 (bottom right).

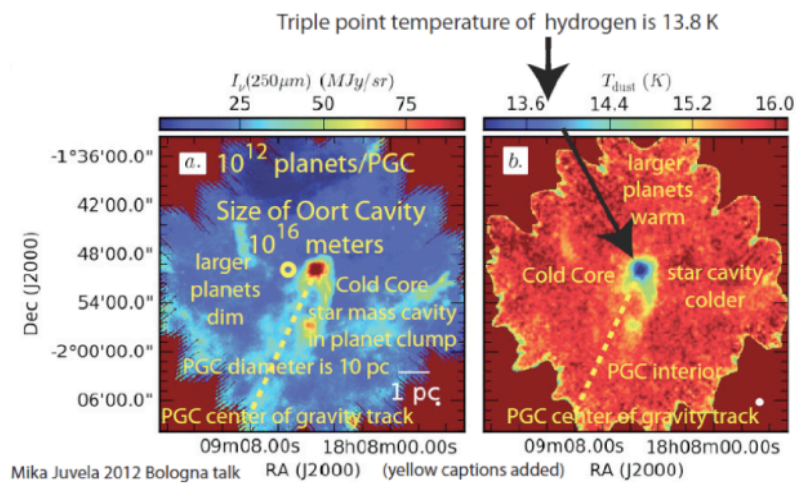

Figure 5. Herschel space telescope image of the infrared radiation (a. left side) and temperature (b. right side) in the interior of a PGC, from Schild, Gibson, Nieuwenhuizen and Wickramasinghe 2012, Fig. 1, based on a 2012 talk by Professor Mika Juvela in Bologna. The dashed yellow lines show the track of another PGC center of gravity, which triggers a Cold Core star cavity to form in the PGC. The weak radiation at high temperatures reflects formation of larger planets by mergers of the $10^{12}$ planets per PGC predicted by HGD cosmology. The strong radiation at low temperatures matching the triple point temperature of hydrogen supports the interpretation of Cold Cores as condensation centers where stars form by frozen hydrogen planet mergers.

As shown in Fig. 5, the infrared radiation and temperature maps of a PGC can be easily understood if these objects are metastable clumps of a trillion dark matter planets, as expected from HGD cosmology. These are identical to giant molecular clouds that are long-term astronomical mystery. They are termed Planck Cold Clouds with Cold Cores. The Cold Core Catalog of Planck Objects is termed C3PO, Juvela et al. 2012, an acronym of the Planck Collaboration.

## 4. Discussion

It seems clear that the dark matter of galaxies is PGC clumps of primordial planets, as shown by Figs. 2, 4 and 5 from the new infrared space telescopes

Herschel, Planck and Spitzer, and others. Several important consequences result. Clearly cold dark matter and dark energy are obsolete. Structure formation starting with the big bang is guided by collisional fluid mechanics. The mechanisms of star formation from frozen gas planets is very different than star formation from gas and dust.

The most important consequence of the $10^{80}$ earth-mass planets emerging from the cosmological big bang is that the provide a clear scenario for complex life to develop, as shown in Figure 6. Life as we know it requires water, but water in abundance will develop on hydrogen planets when seeded by the first stardust by the first supernova events. Supernovae produce oxides of carbon, nitrogen, iron, nickel, silicon etc., all of which will be rapidly reduced to form hot water by the 3000 K hot gas planets emerging from the plasma epoch at 300,000 years.

First the iron and nickel cores of planets will condense, covered by a rocky crust. At 2 million years when the temperature of the universe is near the critical temperature of water 647 K, the first water oceans will begin to condense as critical temperature liquid water, which is apolar and an ideal solvent for the apolar molecules of organic chemistry such as carbon monoxide and hydrocyanic acid shown in Fig. 6.

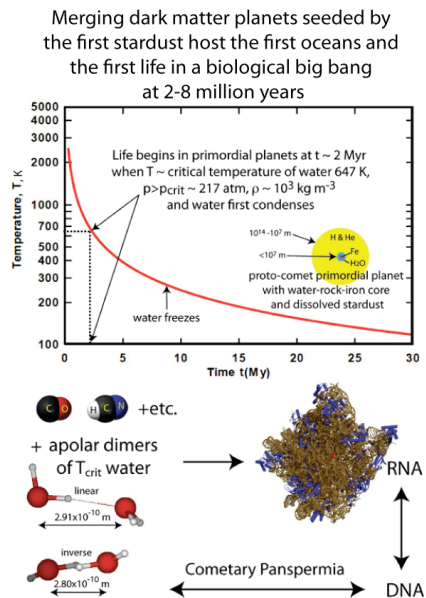

Figure 6. Formation of deep critical temperature water oceans on primordial hydrogen planets that merge to make stars provides a cosmological soup of organic chemical reactions that compete for the carbon produced by stars. The water and genetic codes are spread by supernovae when the stars are overfed by planets and die, confirming the much-maligned Hoyle-Wickramasinghe cometary pansperia hypothesis.

A rather narrow and early window of time between 2 million years when the temperature of the universe reaches 647 K, and 8 million years when the temperature reaches the freezing point of water is referred to as the biological big bang, Gibson, Wickramasinghe and Schild 2010. Life will continue to evolve after the water oceans freeze, but more slowly. The density of the universe at the time of the biological big bang was a million times that at present, and the star formation rate much greater and the oceans hotter. The winning versions of RNA and DNA were very likely formed early in the 2-8 million year window of the biological big bang. The history of its evolution is likely preserved in the DNA of extraterrestrial organisms constantly raining on the Earth.

## 5. Conclusions

Cosmology, astronomy, and astrophysics can be better explained by including modern fluid mechanics, where turbulence is narrowly defined based on the inertial vortex force. Standard collisional fluid mechanics with kinematic viscosity and diffusivity also apply. As shown in Fig. 1, very different time lines result when HGD (hydro-gravitational-dynamics) cosmology is assumed compared to ΛCDMHC (dark-energy cold-dark-matter hierarchical-clustering) cosmology. Fragmentation of the plasma occurs at only 30,000 years, preserving the large baryonic density existing at that time as the density of proto-globular-star-clusters PGCs clumps of a trillion earth-mass gas planets that form the dark matter of galaxies and host the early formation of hot water oceans and primordial life. Infrared space telescopes such as Herschel, Planck, Spitzer etc. now show in great detail the interior of PGCs where stars are formed by planet mergers, Fig. 2 and 5. The transfer of information between hot hydrogen gas planets and their hot water oceans seeded with stardust life nutrients makes the formation and distribution on cosmic scales of RNA and DNA life quite inevitable, as shown in Fig. 6.

The physics of the cosmological big bang event is easily understood by turbulent vortex dynamics and the proposed narrow definition of turbulence based on the inertial vortex force, where the cascade direction of turbulence is from small scales to large, as shown in Fig. 3. It is recommended that the new turbulence and HGD cosmology paradigms be adopted, replacing Taylor-Reynolds-Lumley turbulence and ΛCDMHC cosmology.

## 6. References

Alcock, C. et al., EROS and MACHO combined limits on planetary-mass dark matter in the Galactic Halo, The Astrophysical Journal, 499:L9-L12, 1998.

Ceccarelli, C. and C. Dominik, $H_2D^+$: A light on baryonic dark matter, The Astrophysical Journal, 640: L131-L134, 2006.

Gibson, C. H., Fossil temperature, salinity and vorticity turbulence in the ocean, Marine Turbulence, Proceedings of the 11$^{th}$ International Liege Colloquium on Ocean Hydrodynamics, Jacques C. J. Nihoul, Ed., 1980.

Gibson, C.H., Turbulence in the ocean, atmosphere, galaxy and universe, Appl. Mech. Rev. 49, no. 5, 299–315, 1996.

Gibson, C. H., Turbulence and turbulent mixing in natural fluids, Physica Scripta, Turbulent Mixing and Beyond 2009, T142 (2010) 014030 doi: 10.1088 /0031-8949/2010 /T142/0140302010, arXiv:1005.2772v4, 2010.

Gibson, C. H., Turbulence and Fossil Turbulence lead to Life in the Universe, (Physica Scripta preprint for TMB 2011 Proceedings), Journal of Cosmology, 18, 7951-7963, 2012.

Gibson, C. H., Does cometary panspermia falsify dark energy? Journal of Cosmology, 16, 7000-7003, 2011.

Gibson, C.H. The first turbulence and the first fossil turbulence, Flow, Turbulence and Combustion, 72, 161–179, 2004.

Gibson, C.H. The first turbulent combustion, Combust. Sci. and Tech., 177: 1049–1071, arXiv:astro-ph/0501416, 2005.

Gibson, C. H., N. C. Wickramasinghe, and R. E. Schild, The Biological Big Bang: The First Oceans of Primordial Planets at 2-8 Myr Explains Hoyle – Wickramasinghe Cometary Panspermia and a Primordial LUCA. Journal of Cosmology, 16, 6500-6518, 2011.

Gibson, C. H., R. E. Schild and N. C. Wickramasinghe, The Origin of Life from Primordial Planets, International Journal of Astrobiology 10 (2) : 83–98. 2011.

Gibson, C. H. and R. E. Schild, Evolution of proto-galaxy-clusters to their present form: theory and observations, Journal of Cosmology, 6, 1365-1384, 2010.

Heithausen, A., Molecular Hydrogen as Baryonic Dark Matter, The Astrophysical Journal, 606:L13-L15, 2004.

Hoover, R. B., Fossils of Cyanobacteria in CI1 Carbonaceous Meteorites: Implications to Life on Comets, Europa and Enceladus, Journal of Cosmology, 16, 7070-7111, 2011.

Keating, B. et al., Ultra High Energy Cosmology from POLARBEAR, arXiv:1110.2101v1, 2011.

Pfenniger, D. and F. Combes, Is dark matter in spiral galaxies cold gas? II. Fractal models and star non-formation, Astron. Astrophys. 285, 94-118, 1994.

Pflug, H.D., Heinz, B., Analysis of fossil organic nanostructures: terrestrial and extraterrestrial, Proc. SPIE 3111, 86–97, 1997.

Schild, R. E., Microlensing variability of the gravitationally lensed quasar Q0957+561 A,B. ApJ 464, 125, 1996.

Schild, R. E., C. H. Gibson, Goodness in the Axis of Evil, arXiv:0802.3229v2, 2008.

Walker, M. and M. Wardle, Extreme Scattering Events and Galactic Dark Matter, The Astrophysical Journal Letters, 488 (2), L125, 1998.

Wickramasinghe, N.C., Hoyle, F., Wallis, D.H., Spectroscopic evidence for panspermia, Proc. SPIE 3111, 282–295, 1997.